\documentclass[letter]{aa} 
\usepackage{graphicx}
\usepackage{txfonts}
\usepackage{natbib}
\bibpunct{(}{)}{;}{a}{}{,} 
\begin{document}
\title{AMBER/VLTI observations of the B[e] star MWC 300\thanks{Based on observations made with ESO telescopes at Paranal Observatory under program ID 083.D-0224(C).}}

   \subtitle{}

   \author{Y. Wang
          \inst{1}
          \and
          G. Weigelt\inst{1}
          \and
          A. Kreplin\inst{1}
          \and
          K. -H. Hofmann\inst{1}
          \and
          S. Kraus\inst{2}
          \and
          A. S. Miroshnichenko\inst{3}
          \and
          D. Schertl\inst{1}
          \and
          A. Chelli\inst{4}
          \and
          A. Domiciano de Souza\inst{5}
          \and
          F. Massi\inst{6}
          \and
          S. Robbe-Dubois\inst{5}
          }

   \institute{Max-Planck-Institut f\"ur Radioastronomie, Auf dem H\"ugel 69, 53121 Bonn, Germany\\
   	    \email{ywang@mpifr.de}
         \and
             Department of Astronomy, University of Michigan, 500 Church St., Ann Arbor, MI 48109-1090, USA
         \and
             Department of Physics and Astronomy, University of North Carolina at Greensboro, Greensboro, NC 27402, USA
          \and
             Universit\'{e} Joseph Fourier (UJF) - Grenoble 1\,/\,CNRS-INSU, Institut de Plan\'{e}tologie et d'Astrophysique de Grenoble (IPAG) UMR 5274, Grenoble, F-38041, France
          \and
             Laboratoire Lagrange, UMR 7293, Universit\'{e} de Nice Sophia-Antipolis, CNRS, Observatoire de la C\^{o}te d'Azur, 06300 Nice, France  
         \and
             INAF -- Osservatorio Astrofisico di Arcetri, Largo Fermi 5, 50125 Firenze, Italy       
             }

   \date{}

 
  \abstract
   {}
   {We study the enigmatic B[e] star MWC\,300 to investigate its disk and binary with milli-arcsecond-scale angular resolution.}
   {We observed MWC\,300 with the VLTI/AMBER instrument in the \textit{H} and \textit{K} bands and compared these observations with temperature-gradient models to derive model parameters.
}
   {The measured low visibility values, wavelength dependence of the visibilities, and wavelength dependence of the closure phase directly suggest that MWC\,300 consists of a resolved disk and a close binary. We present a model consisting of a binary and a temperature-gradient disk that is able to reproduce the visibilities, closure phases, and spectral energy distribution. This model allows us to constrain the projected binary separation ($\sim$4.4\,mas or $\sim$7.9\,AU), the flux ratio of the binary components ($\sim$2.2), the disk temperature power-law index, and other parameters.}
   {}


   \titlerunning{AMBER/VLTI observations of the B[e] star MWC\,300}
   \maketitle
%

\section{Introduction}
MWC\,300 is an enigmatic early-type emission-line star whose nature and evolutionary state are not well known. \citet{herbig60} classified MWC\,300 as a pre-main-sequence star since nebulosity exists around the star. It is included in several catalogs of pre-main-sequence stars \citep[e.g.,][]{the94}. However, MWC\,300 was also listed as a probable B[e] supergiant \citep[e.g.,][]{as76, appenzeller77, lamers98}. \citet[][hereafter M04]{miroshnichenko04} carried out extensive spectroscopic and photometric studies on MWC\,300, and strengthened the arguments in favor of the supergiant classification. M04 suggest that MWC\,300 has an effective temperature of 19000\,K, a distance of $1.8\pm0.2$\,kpc, and a luminosity of log\,$L/L_{\odot}=5.1\pm0.1$.

Two-dimensional radiative transfer modeling shows that MWC\,300 has a flared dusty disk rather than a spherical dusty envelope (M04). The modeled disk has an inner radius of $\sim$11\,mas and is viewed almost edge-on. \citet{souza08} observed MWC\,300 at 11.25\,$\mu$m using the BURST mode of VLT/VISIR. The obtained diffraction-limited  ($\sim$0.3$''$) image shows that MWC\,300 is partially resolved with an FWHM diameter of $69\pm10$\,mas. \citet{monnier09} observed MWC\,300 with the Keck telescope and measured a Gaussian FWHM diameter of $49\pm3$\,mas at 10.7\,$\mu$m. 

MWC 300 exhibits radial velocity (RV) variations that suggest a binary \citep{cl99}. \citet{takami03} carried out spectro-astrometric observations and studied the absorption feature of H$\alpha$. They detected a positional displacement of the line-emitting region of $17\pm4$\,mas with respect to the position of the continuum. Photospheric line RV variations detected by M04 also suggest a companion.

\begin{table*}
\caption{Observation log of our VLTI/AMBER observation of MWC\,300.}
\centering
\begin{tabular}{cccccccc}
\hline
\hline
Date & Time & AT Configuration & Projected Baselines & PA & Seeing & DIT & Number of frames\\
 & (UTC) & & (m) & $(^{\circ})$ & ('') & (ms) & \\
 \hline
 2009-04-18 & 07:37 - 07:46 & E0-G0-H0 & 13.90 / 27.78 / 41.68 & 66.2 & 0.5 & 200 & 2400\\
 \hline
 \end{tabular}
 \label{observation}
 \end{table*}
 
 \begin{table*}
 \caption{Scan ranges of the model parameters and their values for the best-fitting models with different inclination angles.}
\centering
\begin{tabular}{cccccccccc}
\hline
\hline
 & $i$\footnotemark[1] & $\theta_{\rm{disk}}$\footnotemark[2] & $r_{\rm{in}}$\footnotemark[3] & $r_{\rm{out}}$\footnotemark[4] & $T_{\rm{in}}$\footnotemark[5] & $\alpha$\footnotemark[6] & $r_{\rm{s}}$\footnotemark[7] & $\kappa$\footnotemark[8] & $\chi^{2}_{\rm{reduced}}$\footnotemark[9] \\
 & ($^{\circ}$) & ($^{\circ}$) & (mas) & (mas) & (K) & & (mas) &  & \\
 \hline
Scan 1 & - & 0--180 & 1--50 & $r_{\rm{in}}$--300 & 500--2000 & 0.0--1.0 & 1--1000 & 1.0--5.0 & -\\
Scan 2 & - & 0--10 & 10--20 & 100--300 & 800--1000 & 0.5--0.7 & 1--20 & 1.0--3.0 & -\\
Results & 80 & 4$\pm3$ & 16.9$^{+1.8}_{-1.5}$ & 200$\pm50$ & 890$\pm20$ & 0.57$^{+0.3}_{-0.1}$ & 4.4$\pm$0.2 & 2.2$^{+0.3}_{-0.2}$ & 2.47 \\
&  & & (30.4$^{+3.2}_{-2.7}$\,AU) & (360$\pm90$\,AU) & & & (7.9$\pm$0.4\,AU) & & \\
 \hline
Scan 1 & - & 0--180 & 1--50 & $r_{\rm{in}}$--300 & 500--2000 & 0.0--1.0 & 1--1000 & 1.0--5.0 & -\\
Scan 2 & - & 10--30 & 5--15 & 100--300 & 800--1000 & 0.5--0.7 & 1--20 & 1.0--3.0 & -\\
Results & 70 & 17$\pm5$ & 11.8$^{+1.0}_{-0.8}$ & 160$\pm40$ & 900$^{+10}_{-20}$ & 0.58$^{+0.2}_{-0.1}$ & 4.4$\pm$0.2 & 2.2$\pm0.2$ & 2.45 \\
&  & & (21.2$^{+1.8}_{-1.4}$\,AU) & (288$\pm72$\,AU) & & & (7.9$\pm$0.4\,AU) & & \\
\hline
Scan 1 & - & 0--180 & 1--50 & $r_{\rm{in}}$--300 & 500--2000 & 0.0--1.0 & 1--1000 & 1.0--5.0 & -\\
Scan 2 & - & 20--40 & 5--15 & 100--300 & 800--1000 & 0.5--0.7 & 1--20 & 1.0--3.0 & -\\
Results & 60 & 29$\pm7$ & 9.7$\pm0.7$ & 140$\pm40$ & 900$\pm10$ & 0.58$\pm0.1$ & 4.4$\pm$0.2 & 2.2$^{+0.2}_{-0.1}$ & 2.44 \\
&  & & (17.5$\pm1.3$\,AU) & (252$\pm72$\,AU) & & & (7.9$\pm$0.4\,AU) & & \\
\hline
 \end{tabular}
 
\footnotemark[1]{Disk inclination angle; }\footnotemark[2]{PA of the long axis of the disk; }\footnotemark[3]{inner disk radius; }\footnotemark[4]{outer disk radius; }\footnotemark[5]{disk temperature at the inner radius; }\footnotemark[6]{temperature power-law index; }\footnotemark[7]{projected binary separation along PA of $66.2^{\circ}$; }\footnotemark[8]{flux ratio between the stars; }\footnotemark[9]{reduced $\chi^{2}$ of the best-fitting model.}
 \label{bestfittingparameter}
\end{table*}

Thanks to the development of near-infrared interferometry, we can now achieve a high spatial resolution of a few mas. It allows us to directly resolve the MWC\,300 system and study the circumstellar disk structure, as well as the central binary. In this paper, we present the first near-infrared interferometric AMBER/VLTI observations of MWC\,300. The observations are described in Section 2. In Section 3, we present a model consisting of a binary and a temperature-gradient disk that can reproduce the visibilities, closure phases, and SED. The results are discussed in Section 4, and conclusions are summarized in Section 5.

\section{Observation and data reduction}

We observed MWC\,300 with the ESO Very Large Telescope Interferometer (VLTI) and the AMBER instrument \citep{petrov07} in the low spectral resolution mode (R = 35) on 18 April 2009 (ID 083.D-0224; PI G. Weigelt). The observation log is presented in Table \ref{observation}. 

For data reduction, we used the AMBER data reduction package \textit{amdlib}\footnote{The AMBER - reduction package \textit{amdlib} is available at: http://www.jmmc.fr/data\_processing\_amber.htm} \citep{tatulli07,chelli09}. To improve the calibration, we used 20\% of the frames with the best fringe signal-to-noise ratio of both the target and calibrator data \citep{tatulli07} and applied a method for equalizing of the target and calibrator histograms of the optical path differences \citep{kreplin12}. The obtained visibilities and closure phases are shown in the Fig. \ref{eomodel} (left two panels). We were not able to derive visibilities at all \textit{H}-band wavelengths (see Fig. \ref{eomodel}) because MWC\,300 is quite faint in the \textit{H} band (\textit{H}=8.2). The data were calibrated using the calibrator HD142669 \citep[diameter = 0.27$\pm$0.027 mas;][]{fracassini01}. 

\section{Temperature-gradient disk model and binary}

The observational data of MWC\,300 shows strong closure phases up to 180$^{\circ}$, which suggest that MWC\,300 is a binary. Therefore, we have used a model consisting of a temperature-gradient disk and a binary to interpret the data.

\subsection{Model description}
Our model consists of a primary star located at the center of the system, a companion, and a ring-shaped disk with a power-law temperature distribution. M04 find that the stellar SED of MWC\,300 can be fitted by a single Kurucz O/B-type model \citep{kurucz94}, and they suggest that the companion of MWC\,300 may be either faint (3--4 mag fainter than the primary) or an O/B-type companion of a similar (but not identical) brightness. We adopt the latter case here, since it is consistent with the detected strong closure phase. We assume that the secondary companion has approximately the same effective temperature as the primary star and that the total flux from the two stars is approximately equivalent to a single star fitted by a Kurucz model with $T_{\rm{eff}}=19000$\,K and $R_{*}=29R_{\odot}$ (M04). The brightness ratio, $\kappa$, between the two stars is a free fitting parameter. Owing to the linear array configuration we used, we can only derive the projected binary separations $r_{\rm{s}}$ along the baseline position angle (PA) of 66.2$^{\circ}$. 

The measured low visibilities suggest that MWC\,300 has a well resolved extended disk.  We assume the disk is geometrically thin and optically thick and that it is emitting as a blackbody. The disk has an inner radius of $r_{\rm{in}}$, outer radius of $r_{\rm{out}}$, and a power-law temperature distribution of $T=T_{\rm{in}}(r/r_{\rm{in}})^{-\alpha}$, where $T_{\rm{in}}$ is the disk temperature at $r_{\rm{in}}$ \citep{hillenbrand92}. Since our interferometric data are measured at only one hour angle with a linear array configuration, we are not able to directly derive the disk elongation and inclination. M04 suggest that the disk of MWC\,300 is viewed almost edge-on, based on their two-dimensional radiative transfer modeling. We adopt their inclination angle of $i=80^{\circ}$ as a fixed value in our modeling.

Our model has seven free parameters (see Table \ref{bestfittingparameter}): the inner disk radius $r_{\rm{in}}$, the outer disk radius $r_{\rm{out}}$, the disk temperature $T_{\rm{in}}$ at the inner radius, the temperature power-law index $\alpha$, the projected separation $r_{\rm{s}}$ of the binary, the flux ratio $\kappa$ between the two stars, and the PA $\theta_{\rm{disk}}$ of the major axis of the inclined disk. The mass center of the binary is expected to be at the center of the disk. However, since we do not know the mass ratio and in order to keep the number of parameters small, we assume that the primary star is located at the center of the disk (for more discussion, see Section 3.2). 

\begin{figure*}
   \centering
   \includegraphics[width=0.9\textwidth]{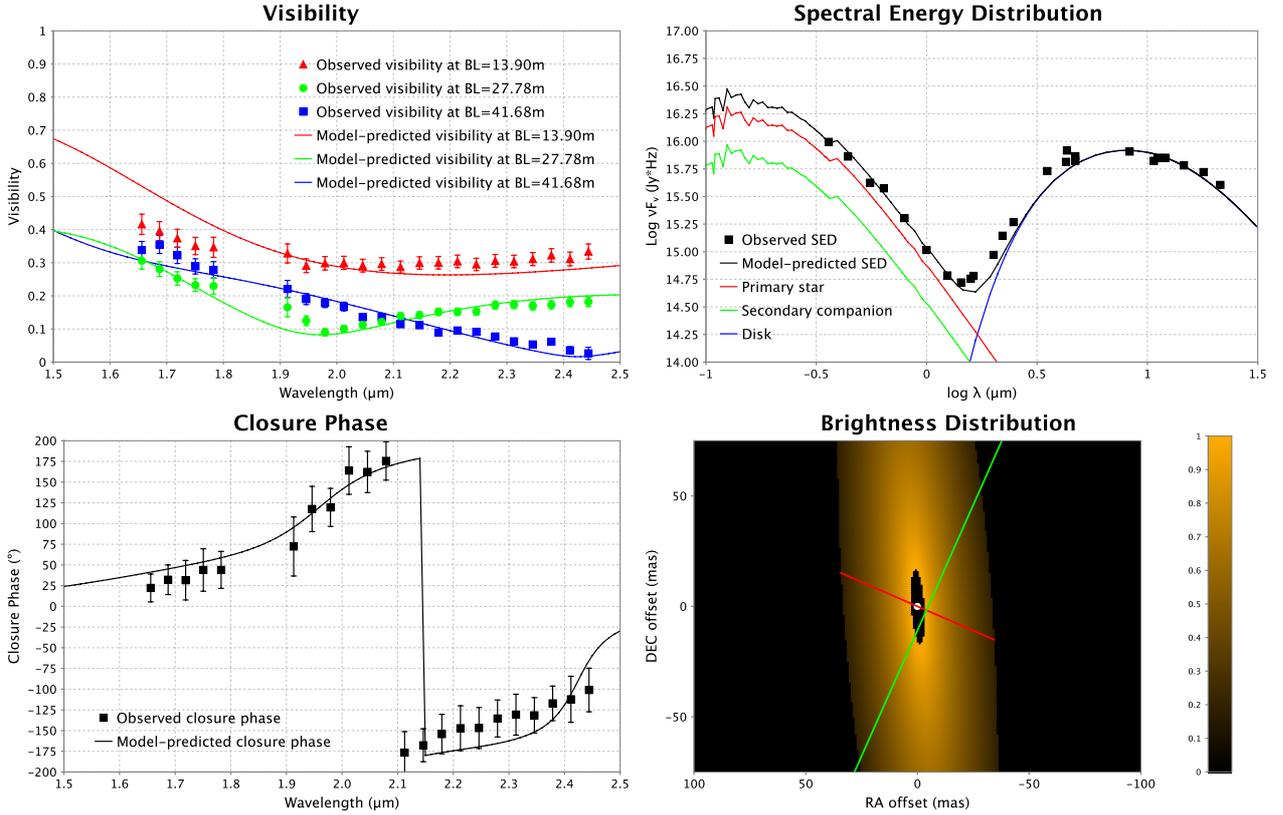}
   \caption{Observations and temperature-gradient model ($i=80^{\circ}$) of MWC\,300. The upper-left panel shows the observed visibilities and the visibilities of the best-fitting model; lower-left the model-predicted and observed closure phases, upper-right the different lines refer to the SED contributions from different components of our model. Black dots are the observed data. The lower-right panel shows the logarithm of brightness distribution of the model disk. North is up and east to the left. The white point represents the location of the primary star. Our observations were performed along the PA of $66.2^{\circ}$ (red line). The green line refers to the possible locations of the secondary companion.}%
   \label{eomodel}
    \end{figure*}

The total complex visibility of our model is given by $\textbf{V}_{\rm{total}}=f_{\rm{primary}}\textbf{V}_{\rm{primary}}+f_{\rm{secondary}}\textbf{V}_{\rm{secondary}}+f_{\rm{disk}}\textbf{V}_{\rm{disk}}$, where $f$ is the flux contribution of each component to the total system flux and $\textbf{V}$ the complex visibility contributed by each component. We calculated the model-predicted visibilities and closure phases for each model (Table 2; Section 3.2) and compared them with the observations in order to study if our model can simultaneously reproduce all interferometric observations, as well as the spectral energy distribution (SED). The dereddened SED data (shown in the upper right panel in Fig. \ref{eomodel}) is taken from M04 and \citet{souza08}, including $UBV$ data from \citet{ws85}, $JHK$ data from \citet{skrutskie06}, $LM$ data from \citet{bs82}, $NQ_{s}$ data from M04, and MSX B1B2ACDE data from \citet{egan03}.

\subsection{Results of the temperature-gradient modeling}

We searched for the best-fitting model in a large parameter space (see two-step search in Table~\ref{bestfittingparameter}). During each step, we divided the search range for each of the model parameters into 10--20 intervals and calculated the visibilities, closure phases, and SEDs of all models that correspond to all combinations of all free model parameter values ($\sim$$2\times8\times10^{7}$ models; see Table \ref{bestfittingparameter}). Then the derived visibilities, closure phases, and SEDs of all $1.6\times10^{8}$ models were compared with the observations in order to find the best-fitting model with the minimum overall reduced $\chi^{2}$.

Figure \ref{eomodel} and Table \ref{bestfittingparameter} show our best-fitting MWC\,300 models, which are able to approximately reproduce the observed wavelength dependence of the visibilities and the closure phase, along with the SED. In our first model (Fig. \ref{eomodel} and first three lines in Table \ref{bestfittingparameter}), we adopted the inclination angle of $i=80^{\circ}$ (M04) as a fixed value, as discussed above. The flux contributions from the binary and the disk at different wavelengths (e.g. 1.6\,$\mu m$, 1.9\,$\mu m$, 2.1\,$\mu m$, and 2.4\,$\mu m$) are listed in Table~\ref{fluxratio}. This modeling provided the following parameters. The secondary companion is $2.2^{+0.3}_{-0.2}$ times fainter than the primary star and has a projected separation of $4.4\pm0.2$\,mas or $7.9\pm0.4$\,AU along the PA of $66.2^{\circ}$. The temperature-gradient disk has an inner radius of $\sim$17\,mas or $\sim$30\,AU and an outer radius $r_{\rm{out}}$ of $\sim$200\,mas or $\sim$360\,AU, while $r_{\rm{out}}$ cannot be well constrained since the outer (colder) parts of the disk are faint in the infrared. The disk major axis has a PA of $\sim$4$^{\circ}$. The temperature at the inner disk radius is $\sim$890\,K, and the temperature power-law index is $\sim$0.57 (see Table \ref{bestfittingparameter}). All uncertainties correspond to the 5$\sigma$ confidence level.
    
We also calculated models assuming that the geometric center of the binary instead of the primary star is at the center of the disk. However, these models cannot provide a better fit. Therefore, in the following, we only discuss the models with the primary star located at the center of disk.

\section{Discussion}

\subsection{Disk inclination angle}
\label{diskinclination}

The emission line profile of MWC\,300 and strong extinction along the line of sight suggest an almost edge-on disk (M04). The value of 80$^{\circ}$ (i.e., almost edge-on) derived by  radiative transfer modeling (M04) is only a rough estimate, so we also investigated models with other inclination angles. The values of parameters for the best-fitting models with $i=60^{\circ}$ and $70^{\circ}$ are listed in Tabel \ref{bestfittingparameter} (lines 4-9). These models can fit the observations with similar $\chi^{2}$ values. Our results in Table \ref{bestfittingparameter} show that some of our model parameters (e.g., $T_{\rm{in}}$, $\alpha$, $r_{\rm{s}}$, and $\kappa$) are independent of the disk inclination angle. However, $r_{\rm{in}}$ and $\theta_{\rm{disk}}$ strongly depend on the disk inclination angle and cannot be constrained well. Our interferometric data do not yet allow us to constrain the inclination angle since our three visibilities were measured at the same PA. Therefore, more observations at different PAs are required in the future to improve the accuracy of the disk parameters.

\subsection{Disk size}
M04 modeled MWC\,300 with a two-dimensional flared dusty-disk model and suggest that the inner boundary of the disk has a radius of $\sim20$\,AU and a temperature of $\sim$1120\,K. With the same assumed disk inclination angle of 80$^{\circ}$, our best-fitting model suggests a slightly larger and cooler disk, with an inner radius of $\sim$30\,AU and a temperature at the inner radius of $\sim$890\,K. This radius is consistent with the dust sublimation radius of $\sim$35\,AU for 890\,K and gray dust opacities, but is about 2.5 times larger than the 1500\,K dust sublimation radius of $\sim$12\,AU \citep{mm02}. This large inner disk radius and cool inner disk temperature may be explained by the existence of the secondary companion. Considering the minimum binary separation of 4.4\,mas (7.9\,AU), MWC\,300 may have a circumbinary disk, and the secondary companion may be close enough to the disk inner radius to interact strongly with the disk \citep{shi12}.

\begin{table}
\caption{Flux contribution from different components at different wavelengths.}
\centering
\begin{tabular}{ccccc}
\hline
\hline
 & 1.6\,$\mu m$ & 1.9\,$\mu m$ & 2.1\,$\mu m$ & 2.4\,$\mu m$ \\
 \hline
$f_{\rm{primary}}$ & 0.491& 0.229& 0.124& 0.052\\
$f_{\rm{secondary}}$ & 0.223& 0.104 & 0.056& 0.024\\
$f_{\rm{disk}}$ & 0.286 & 0.667 & 0.82& 0.924\\
 \hline
 \end{tabular}
 \label{fluxratio}
 \end{table}

\subsection{Binarity and B[e] phenomenon}

Our interferometric data shows strong closure phases that suggest asymmetries. The asymmetries can be attributed to a binary or an extremely inhomogeneous disk. However, the observed closure phase signature is too strong to be caused by the asymmetry of the disk since the disk is only partially resolved \citep{monnier06}. The observed closure phase is therefore the result of a secondary companion.

Adopting the distance of 1.8\,kpc, we obtain a small projected binary separation of $\sim$7.9\,AU. However, due to our one-dimensional $uv$ coverage, we can only derive the projected separation of the binary along PA of $66.2^{\circ}$ and not its exact separation. The lower-right hand panel of Fig. \ref{eomodel} indicates the possible locations of the secondary companion. All companion positions on the green line correspond to a projected binary separation of 7.9\,AU. The figure shows that MWC\,300 may have a circumbinary disk if the binary separation is small enough \citep[as, for example, in HD\,62623; ][]{plets95,millour09}. However, considering the AMBER AT's field of view, the binary separation can be large enough (with a maximum value of 150\,mas) for the disk to be a circumprimary disk \citep[as, for example, in HD\,87643; ][]{millour09}. More interferometric observations are required in order to constrain the exact separation and PA of the binary. 

Several other B[e] stars were also found to be binaries: $\eta$ Car, MWC\,349A, HD\,87643, V921\,Sco, and HD\,327083. In HD\,87643 \citep{millour09} and V921\,Sco \citep{kraus12a}, circumprimary disks were resolved, whereas in HD\,327083 \citep{wheelwright12} a circumbinary disk was detected. The $\eta$ Car binary was found by photometric observations \citep{damineli97}. The binarity of MWC\,349A has not been directly confirmed, but was suggested by \citet{hofmann02}. Although the exact role of binarity in the B[e] phenomenon is still an open question, it is possible that the disks are the result of mass exchange episodes between the stellar components of the B[e] stars \citep{sheikina00, miroshnichenko07a, miroshnichenko07b, miroshnichenko09, kraus10, kraus12b}. The slowing down of the radiative wind by the companion gravitational effect or wind shock may also be an explanation \citep{plets95}. Future interferometric observations will allow us to improve our understanding, since these observations will be able to determine the orbits of the B[e] binaries and to investigate the interaction of the binary companion with the disk or the wind.

\section{Summary}

We have presented the first VLTI/AMBER observations of the enigmatic B[e] star MWC 300. The low visibility values, the wavelength dependence of the visibilities, and the wavelength dependence of the closure phase directly suggest that MWC\,300 consists of a resolved disk and a binary. To interpret our observations, we used a temperature-gradient model consisting of a binary and a ring-shaped disk. The best-fitting model is able to simultaneously reproduce the wavelength dependence of the visibilities and the closure phase, as well as the SED. Because of our one-dimensional $uv$ coverage, we can only derive a projected binary separation of $\sim$4.4\,mas ($\sim$7.9\,AU) along PA of 66.2$^{\circ}$, with a binary flux ratio of $\sim$2.2 (see Table 2).  The increasing number of detected binaries in B[e] stars suggests that the binarity may play an important role in the B[e] phenomenon.

\begin{acknowledgements}
      We thank the ESO VLTI team on Paranal for the excellent collaboration. The data presented here were reduced using the publicly available data-reduction software package \textit{amdlib} kindly provided by the Jean-Marie Mariotti Center. 
\end{acknowledgements}

\bibliographystyle{aa} 
\bibliography{wyYSO} 

\end{document}